\pdfoutput=1	
\documentclass[9pt,twocolumn,twoside]{pnas-new}
\setboolean{displaywatermark}{false}

\templatetype{pnasresearcharticle}

\title{Electron-Tunnelling-Noise Programmable Random Variate Accelerator for Monte Carlo Sampling}

\author[1]{James T. Meech}
\author[2]{Vasileios Tsoutsouras}
\author[1,2]{Phillip Stanley-Marbell} 

\affil[1]{Department of Engineering, University of Cambridge}
\affil[2]{Signaloid}
\leadauthor{James T. Meech} 

\authorcontributions{}

\vspace{-0.1in}

\usepackage{framed}
\usepackage{moreverb}
\usepackage{xspace}
\usepackage{flushend}			%

\usepackage{xfrac}			%

\usepackage{amsfonts}
\usepackage{amssymb}    
\usepackage{amsmath}
\usepackage[thmmarks,amsmath,hyperref]{ntheorem}
\usepackage{graphicx}
\usepackage{times}
\usepackage[scaled]{helvet} 		%
\usepackage{textcomp}
\usepackage{color}
\usepackage{colortbl}
\usepackage{booktabs}
\usepackage{listings}
\usepackage{multicol}
\usepackage{times}
\usepackage{pifont}			%
\usepackage{microtype}
\usepackage{siunitx}
\usepackage[ruled,vlined,linesnumbered]{algorithm2e}
\usepackage{subcaption}

\definecolor{a}{rgb}{0.83, 0.83, 0.83}
\definecolor{b}{rgb}{0.99,0.99,0.99}
\definecolor{y}{rgb}{1.0, 0.94, 0.0}

\usepackage{mmm}
\usepackage{wand-sample}
\usepackage[scaled=0.85]{beramono}
\usepackage[T1]{fontenc}
\usepackage{url}
\usepackage{multirow}
\usepackage{array}

\definecolor{listinggreen}{rgb}{0,0.6,0}
\definecolor{listinggray}{rgb}{0.5,0.5,0.5}
\definecolor{listingmauve}{rgb}{0.58,0,0.82}
\definecolor{listingkeywordcolor}{rgb}{1.0,0.4,0.0}
\definecolor{listinglightgray}{rgb}{0.9863,0.9863,0.9863}

\lstdefinelanguage{FSharp}
{
	morekeywords	= {
    let,
    type,
    Measure,
	},
	sensitive	= false,
	morecomment	= [l]{\#},
	morecomment	= [s]{(*}{*)},
}

\lstdefinelanguage{Newton}
{
	morekeywords	= {
		signal,
		derivation,
		symbol,
		name,
		invariant,
		constant,
		English,
		sensor,
		name,
		none,
		dot,
		cross,
		derivative,
		integral,
		interface,
		i2c,
		spi,
		analog,
		write,
		read,
		delay,
		range,
		erasuretoken,
		uncertainty,
		accuracy,
		precision,
		Gaussian,
		exponential,
		biexponential,
		to,
		bits,
		dimensionless,
		include
	},
	sensitive	= false,
	morecomment	= [l]{\#},
	morecomment	= [s]{/*}{*/},
}

\usepackage{hyperref}
\hypersetup{
    colorlinks=false, %
    linktoc=all,     %
    linkcolor=blue,  %
}

\begin{abstract}
This article presents an electron tunneling noise programmable random variate accelerator for accelerating the sampling stage of Monte Carlo simulations.
We used the LiteX framework to generate a FemtoRV \texttt{imfc} RISC-V instruction set soft processor and deploy it on a Digilent Arty-100T FPGA development board.
The RISC-V soft processor augmented with our programmable random variate accelerator achieves an average speedup of $8.70 \times$ and a median speedup of $8.68 \times$ for a suite of twelve different benchmark applications when compared to GNU Scientific Library software random number generation. 
These speedups are achievable because the benchmarks spend an average of 90.0\,\% of their execution time generating random samples. 
The results of the Monte Carlo benchmark programs run over the programmable random variate accelerator have an average Wasserstein distance of $1.48 \times$ and a median Wasserstein distance of $1.41 \times$ that of the results produced by the GNU Scientific Library random number generators.
The soft processor samples the electron tunneling noise source using the hardened XADC block in the FPGA.
The flexibility of the LiteX framework allows for the deployment of any LiteX-supported soft processor with an electron tunneling noise programmable random variate accelerator on any LiteX-supported development board that contains an FPGA with an XADC.
\end{abstract}

\dates{This manuscript was compiled on \today}

\begin{document}
\verticaladjustment{-5pt}

\maketitle
\thispagestyle{firststyle}
\ifthenelse{\boolean{shortarticle}}{\ifthenelse{\boolean{singlecolumn}}{\abscontentformatted}{\abscontent}}{}

\section{Introduction}\label{section:introduction}
\vspace{-0.1in}
\dropcap{M}{onte} Carlo simulations for quantifying uncertainty in computations are becoming increasingly more important~\cite{tsoutsouras2021laplace,tsoutsouras2022laplace}. 
Computer systems are expected to make increasingly complicated real-time decisions in life-critical applications using uncertain data~\cite{choi2019gaussian}. 
At the same time, Moore's law has ceased to provide increased computer hardware performance through transistor scaling laws~\cite{tye2023materials}. 
These two events call for domain-specific architectures to accelerate applications such as Monte Carlo simulations~\cite{dally2023domain, hennessy2019new}.
This article introduces a domain-specific hardware accelerator for the sampling stage of Monte Carlo simulations.
Our accelerator will complement other domain-specific architectures designed for propagating uncertainty through calculations~\cite{tsoutsouras2021laplace,tsoutsouras2022laplace}.
The programmable random variate accelerator is a faster and more efficient replacement for digital electronic hardware for Monte Carlo sampling.
The programmable random variate accelerator replaces function calls to random number generator libraries with code to sample from the programmable random variate accelerator.
The accelerator uses an analog-to-digital converter with direct memory access to allow a RISC-V instruction set processor to sample from any univariate probability distribution using a physics-based programmable non-uniform random variate generator. 

\subsection{Contributions}

This article presents the following contributions:

\begin{enumerate}
	\item \textit{A programmable random variate accelerator design that can sample from arbitrary univariate Gaussian distributions and arbitrary univariate distributions as Gaussian mixtures.}
	\item \textit{A prototype interfacing the programmable random variate accelerator with a FemtoRV Petitbateau \texttt{imfc} RISC-V instruction set soft processor on a field programmable gate array.}
	\item \textit{Evaluation of the speed and efficiency improvements gained by using the programmable random variate accelerator to accelerate a suite of twelve benchmark applications.}
	\item \textit{Comparison of the programmable random variate accelerator Monte Carlo simulation accuracy for a suite of twelve benchmark applications using Wasserstein distance from a reference result.}
\end{enumerate}

\subsection{Motivation for Monte Carlo Sampling}

\begin{figure*}
	\centering
	\includegraphics[width=\textwidth]{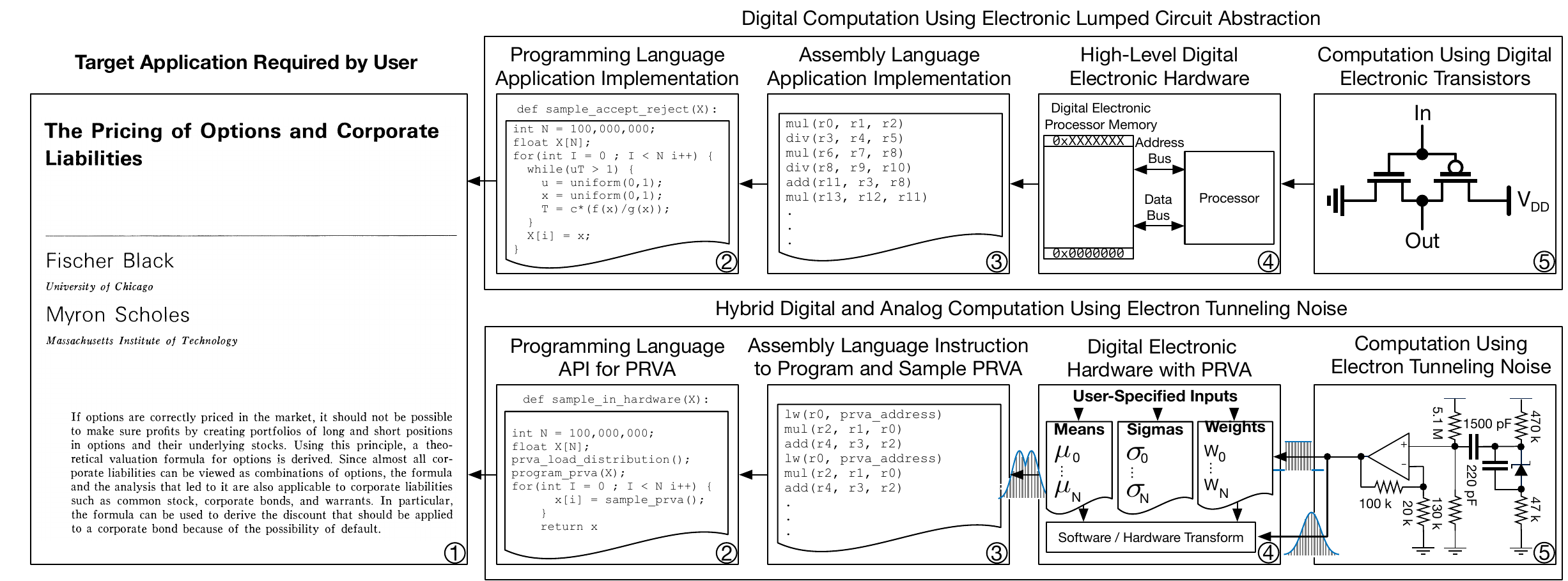}
	\caption{The steps required to perform a Black Scholes Monte Carlo simulation using a programmable random variate accelerator instead of a digital electronic processor diverge at the first abstraction level below the abstract Black Scholes application required by the user. 
	         The programmable random variate accelerator requires small changes at every level of the software and hardware stack to use electron tunneling noise to generate samples from parameterized probability distributions to run the Monte Carlo simulation.}\label{fig:bigPicture}
\end{figure*}

Figure~\ref{fig:bigPicture} panel~\ding{192} shows an application where the user wants to generate random samples to perform a Black Scholes Monte Carlo simulation. 
The application code will call a random number generation function that will perform the Box-Muller transform on the output of a uniform pseudorandom number generator to obtain the samples from a Gaussian distribution required for the Black-Scholes method.
Traditional digital electronic hardware will then execute assembly instructions to run the program for the user.
For named probability distributions with a closed-form inverse cumulative distribution function other than the Gaussian the random number generation function will use the inversion method.
If there is no closed form for the inverse cumulative distribution function the random number generation function will resort to using an even more inefficient rejection sampling method~\cite{devroye1986nonuniform}.
All of the math operations required for the application will be converted to assembly instructions to be run on an in-order digital electronic processor. 
In this architecture, each instruction and the corresponding data will have to be read from the main memory over the data bus incurring long latency. 
Finally, the digital electronic transistors that implement the hardware will switch the flow of electrons to perform the computation.

A more efficient alternative to the traditional hardware shown in the top panel of Figure~\ref{fig:bigPicture} would be to offload all of the random number generation to a dedicated programmable random variate accelerator. 
The user can simply replace all random number generator calls with calls to the programmable random variate accelerator application programming interface.
Then the compiler generates assembly instructions to program and sample from the programmable random variate accelerator. 
These assembly instructions will allow the programmable random variate accelerator to use an analog-to-digital converter to sample from a Gaussian fast and efficiently using an analog noise source. 
The programmable random variate accelerator will then quickly and efficiently transform the raw Gaussian samples to samples from the required probability distribution and store them in a register in the processor. 
This replaces the multiple assembly instructions required to perform the Box-Muller transform, inversion, or rejection sampling by a single instruction to sample from the programmable random variate accelerator. 

\section{Programmable Random Variate Acceleration}

Sampling from empirical application-specific non-uniform probability distributions is slow and inefficient. 
Digital electronics processors use the fast and efficient inversion method~\cite{devroye1986nonuniform} shown in Algorithm~\ref{algorithm:inversion} to transform uniformly-distributed random samples to non-uniformly-distributed random samples.
Digital electronic processors can only use the inversion method to obtain random samples for an application when there exists an analytical closed-form solution for the inverse cumulative distribution function of the probability density function that the application requires samples from.
For probability density functions for which there is no analytic form for the inverse cumulative distribution function,
the digital electronic processor must instead use the accept-reject method~\cite{devroye1986nonuniform} to transform random samples with a uniform distribution to samples from the target non-uniform distribution. 
Algorithm~\ref{algorithm:accept-reject} shows the accept-reject method. 
The accept-reject method repeatedly generates samples until a sample satisfies the condition on line four of Algorithm~\ref{algorithm:accept-reject}.
The algorithm discards any samples that do not meet the condition.
As a result, the accept-reject method is slow and inefficient. 

\begin{algorithm}
\DontPrintSemicolon
\SetAlgoLined
\KwResult{Sample from non-uniform random variable $X$}
Generate uniform [0, 1] random variate $u$\;
RETURN $x \leftarrow F^{-1}(u)$\;
\caption{Inversion method~\cite{devroye1986nonuniform}. 
Let $X$ be a non-uniform random variable that the digital electronic processor needs to sample, 
$U$ be the uniform random variable that the digital electronic processor can sample from, 
and $F^{-1}$ the inverse cumulative distribution function that the digital electronic processor requires samples from.}\label{algorithm:inversion}
\end{algorithm}

\begin{algorithm}
\DontPrintSemicolon
\SetAlgoLined
\KwResult{Sample from non-uniform random variable $X$}
\Repeat{$uT \leq 1$}{
Generate uniform [0, 1] random variate $u$\;
Generate uniform [0, 1] random variate $x$\;
Set $T \leftarrow c \frac{f(x)}{g(x)}$\;
}
RETURN $x$\;
\caption{Accept-reject method~\cite{devroye1986nonuniform}. 
Let $g$ be the density of $U$ and $f$ be the density of $X$. 
Let $c \geq 1$ be a constant such that the condition $f(x) \leq cg(x)$ holds for all $x$.
The accept-reject method probabilistically rejects random variates from a probability distribution that is easy to sample to produce samples from a probability distribution that is hard to sample.}\label{algorithm:accept-reject}
\end{algorithm}

\section{Arbitrary Distribution From Gaussian Mixture}

Many Monte Carlo simulations require samples from bespoke empirical non-uniform probability distributions.
Our analog noise source can only generate samples from a univariate Gaussian.
We therefore present the theory required to generate samples from any univariate probability distribution using samples from a univariate Gaussian random variable.
Starting from a univariate distribution described in terms of discrete samples, it is possible to reconstruct the probability density function using a kernel density.
In the case where we select the Gaussian as our kernel function the kernel density is a mixture of Gaussian distributions. 

A programmable random variate accelerator can generate samples from a Gaussian distribution using Gaussian electronic noise and then transform those samples to change the mean and standard deviation of the distribution.
The programmable random variate accelerator can generate samples from any non-uniform distribution by decomposing it into a mixture of Gaussians and then using the Gaussian-to-Gaussian transform (Section~\ref{section:gaussiantogaussian}) to sample from each Gaussian component of the kernel density. 
The accelerator can then use a uniform-random-number generator to sample from each Gaussian distribution with likelihood proportional to the relative weight of each Gaussian (or use equal weights).  

\subsection{Kernel Densities} 

If we take a set of $N$ data points that represent a distribution, we can construct a kernel density estimate of the probability density function. 
We do this by fitting a Gaussian mixture to the points. 
Let $N$ be the number of samples, $M$ be the number of component functions, $K$ be the kernel density function that we use (a Gaussian distribution) and $h$ be the bandwidth which is a smoothing parameter.
Wand et al.~\cite{wand1994kernel} show that we can write the normalized version of our kernel density estimate as 

\begin{equation}
\hat{f}_X(x) = \frac{1}{Mh} \sum^M_{i=1} K \bigg( \frac{x-x_i}{h} \bigg).
\label{equation:kernelDensity}
\end{equation}
 
Assuming that the data has a Gaussian distribution we can use Silverman's rule to calculate $h$. 
Let $\sigma_N$ be the standard deviation of all $N$ data points,
Silverman~\cite{silverman1986density} shows that we can estimate $h$ as 

\begin{equation}
h = \bigg( \frac{4 \sigma_N^5}{3N} \bigg)^{\frac{1}{5}}.
\label{equation:silvermansRuleOfThumb}
\end{equation}

More sophisticated kernel density estimation methods exist if the application demands a better approximation~\cite{heidenreich2013bandwidth, chiu1996comparative, sheather1991reliable}.

\subsection{Gaussian-to-Gaussian Transform}\label{section:gaussiantogaussian}

The programmable random variate accelerator can obtain a Gaussian-distributed random variable $X'$ with any desired mean $\mu'$ and standard deviation $\sigma'$
from a Gaussian-distributed random variable $X$ with a mean $\mu$ and standard deviation $\sigma$~\cite{leemis2008univariate} by applying the transformation

\begin{equation}
    X' = (Xa) + b 
\label{equation:normalToNormalTransform} 
\end{equation} 
where 

\begin{equation}
    a = \frac{\sigma'}{\sigma}
\label{equation:normalToNormalTransforma} 
\end{equation}  
and
\begin{equation}
    b = \mu' - \mu a.
\label{equation:normalToNormalTransformb} 
\end{equation}  

Sampling from a random variable by transforming samples from an existing Gaussian or mixture of Gaussian random variables always produces a sample, unlike the accept-reject method which does not. 

\section{Electron Tunelling Noise Source Implementation}\label{section:implimentation}
The implementation of the Gaussian random number generator consists of a constant current reverse bias generating circuit, a reverse-biased Zener diode noise source, a direct current blocking capacitor, a bias setting potential divider, and an operational amplifier.
Both the reverse bias generating circuit and the oper can be turned off to reduce power consumption by connecting the ``On'' pin to 0\,V.

\subsection{Circuit Design}\label{section:circuitDesign}
\begin{figure}
\centering
\includegraphics[width=0.485\textwidth]{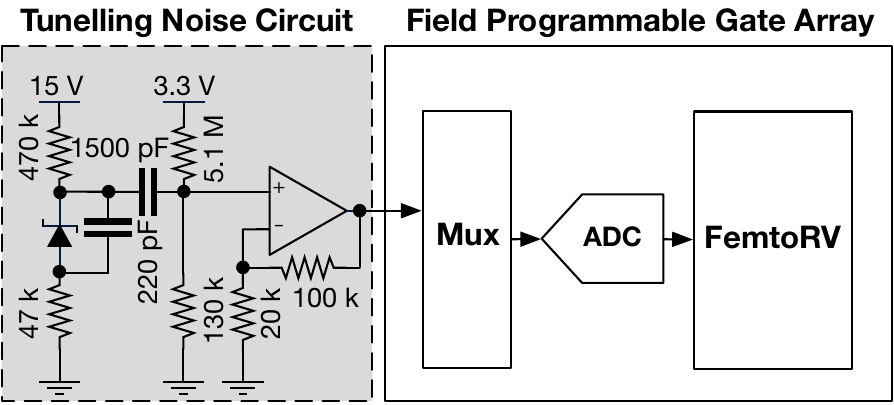}
\caption{The physics-based Gaussian random number generator circuit.
		 Details of circuits such as the constant current light emitting diode driver circuit that produces the 15\,V bias voltage are omitted here but available in~\cite{avalanche-noise}.}\label{fig:circuitDiagram}
\end{figure}

\begin{figure}
	\centering
	\includegraphics[width=0.485\textwidth]{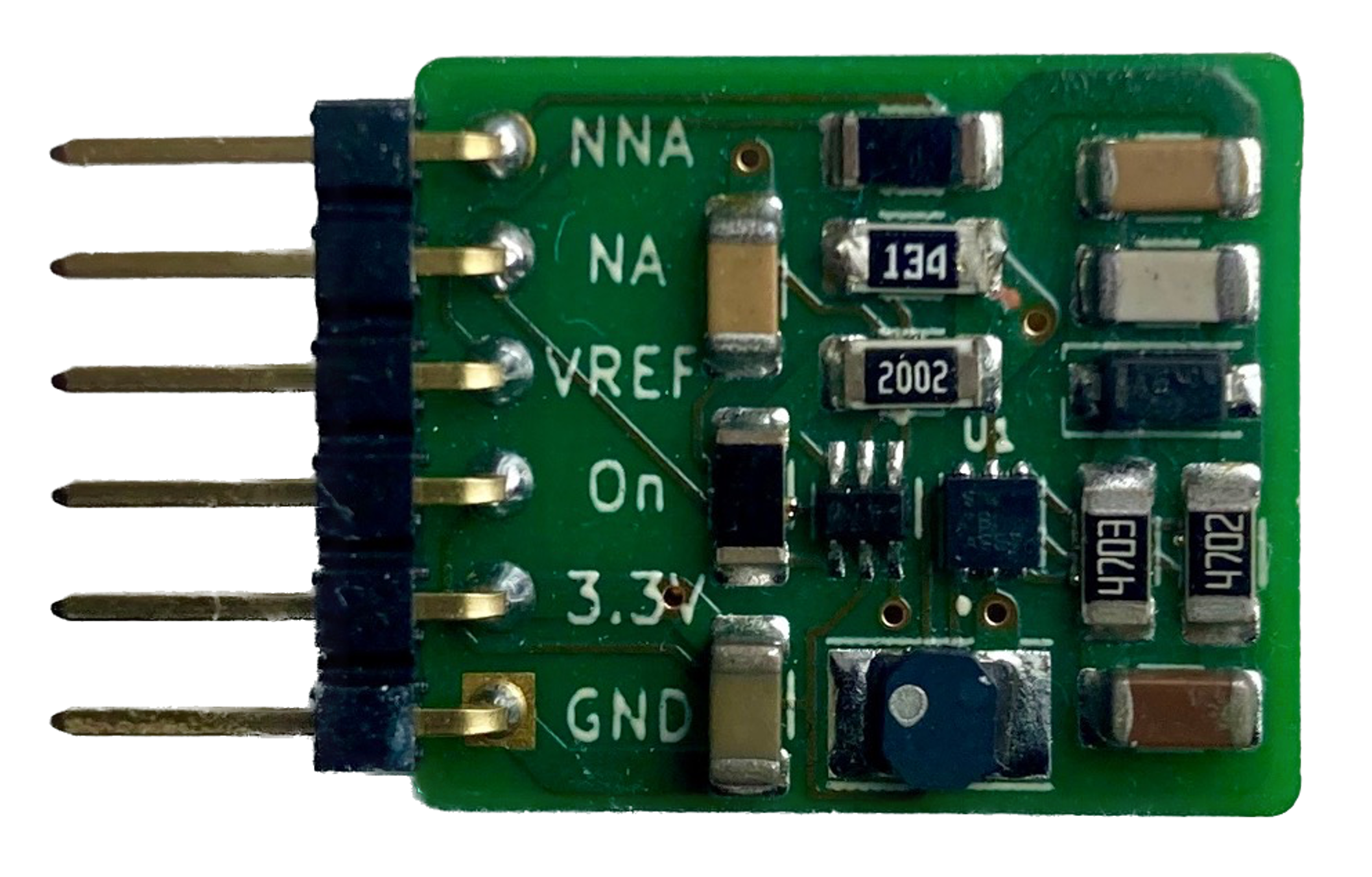}
	\caption{The low power Gaussian noise source printed circuit board.
			 The printed circuit board consumes 1.62\,mW of power in the on state and 32.4\,nW in the off state~\cite{avalanche-noise}.}\label{fig:printedCircuitBoard}
\end{figure}

Figure~\ref{fig:circuitDiagram} shows the circuit diagram of the programmable noise source design slightly modified from prior work by Huang~\cite{avalanche-noise}.
The design leverages a high voltage constant current light emitting diode driver to provide a Zener diode with a 15\,V reverse bias.
The light-emitting diode driver enforces a constant current to minimize the overall power consumption. 
The reverse-biased Zener diode generates a noise voltage which is amplified by an operational amplifier with a gain of $5\times$.
Prior to this, a direct current blocking capacitor removes the large 15\,V bias voltage from the noise signal and a potential divider sets the mean of the noise signal.
Figure~\ref{fig:printedCircuitBoard} shows a prototype printed circuit board that implements the circuit shown in the shaded gray box on the left side of Figure~\ref{fig:circuitDiagram} (i.e., excluding the mux, ADC, and CPU).

The circuit uses a constant current light emitting diode driver powered by the supply rail to apply a 15\,V reverse bias to the Zener diode to cause electrons to randomly tunnel through the PN junction. 
This random tunneling activity manifests itself as a random voltage signal that the circuit amplifies using a non-inverting operational amplifier circuit.
The direct current blocking capacitor removes the 15\,V direct current bias, and the circuit adds a new bias to the signal using a programmable resistor divider. 
The resistor divider controls the mean of the Gaussian noise distribution.
The gain of the non-inverting operational amplifier controls the standard deviation of the distribution. 
Finally, the analog-to-digital converter quantizes the output of the amplifier to 12-bit unsigned integers.
We set the bias and gain to ensure that the analog-to-digital converter sees a large range of unique signal values and captures the Gaussian shape of the distribution.

\begin{figure}
\centering
\includegraphics[width=0.485\textwidth]{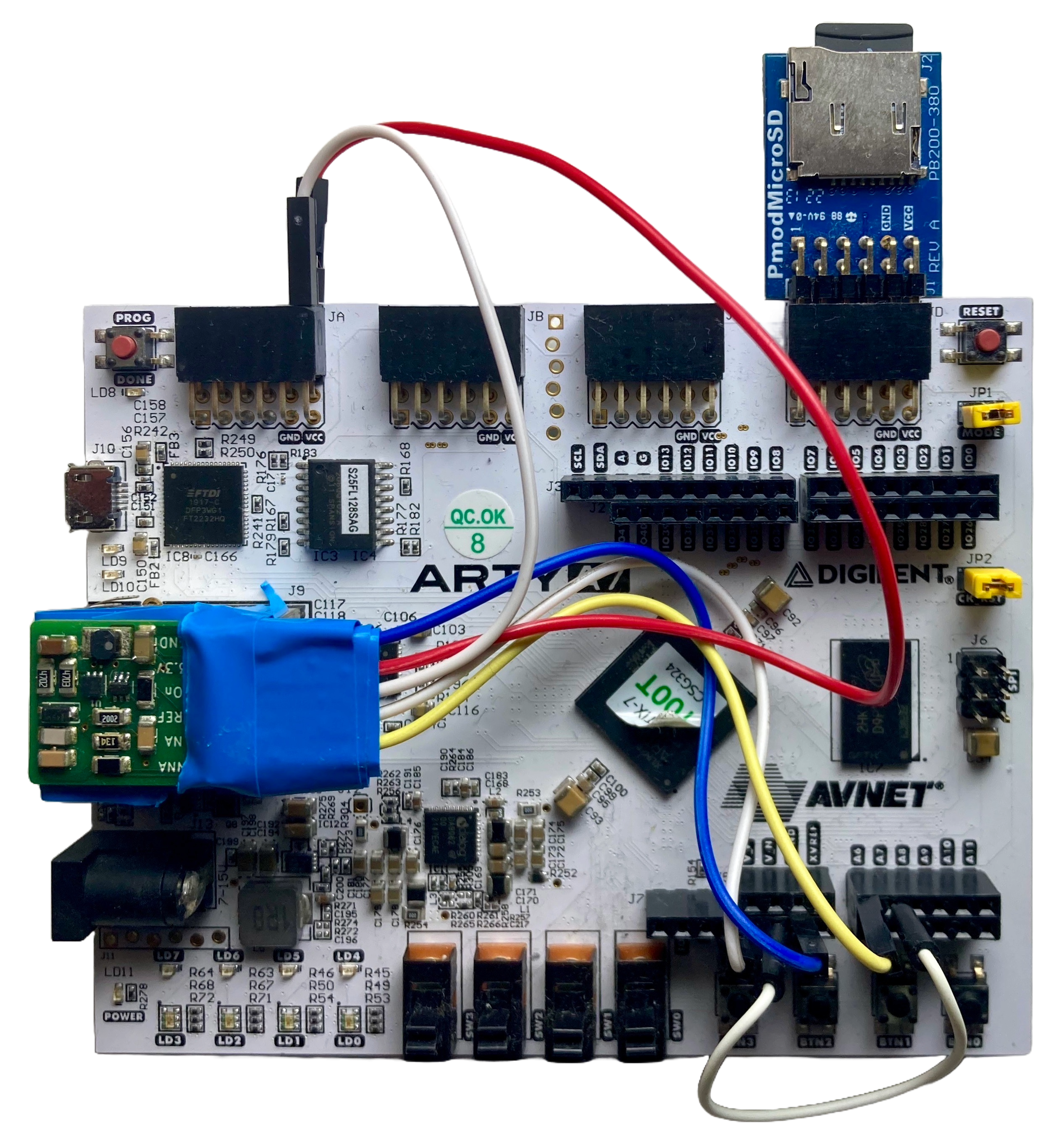}
\caption{The low power Gaussian noise source printed circuit board connected to the XADC of an Artix-7 XC7A100T FPGA on a Digilent Arty development board.
		 The microSD card PMOD stores .elf program files to be executed on the soft processor on the FPGA.
		 LiteOS provides the functionality to compile C language programs to .elf files for the RISC-V Petitbateau soft processor~\cite{liteos}. 
		 The FPGA development board and noise source board combined consume approximately 1.983\,W of power when sampling from a univariate Gaussian.}\label{fig:developmentBoardFPGA}
\end{figure}

\subsection{Kernel Density Programmable Random Variate Accelerator for Programmable Univariate Distributions}\label{section:kernelPRVA}
\begin{figure}
	\centering
	\includegraphics[width=0.485\textwidth]{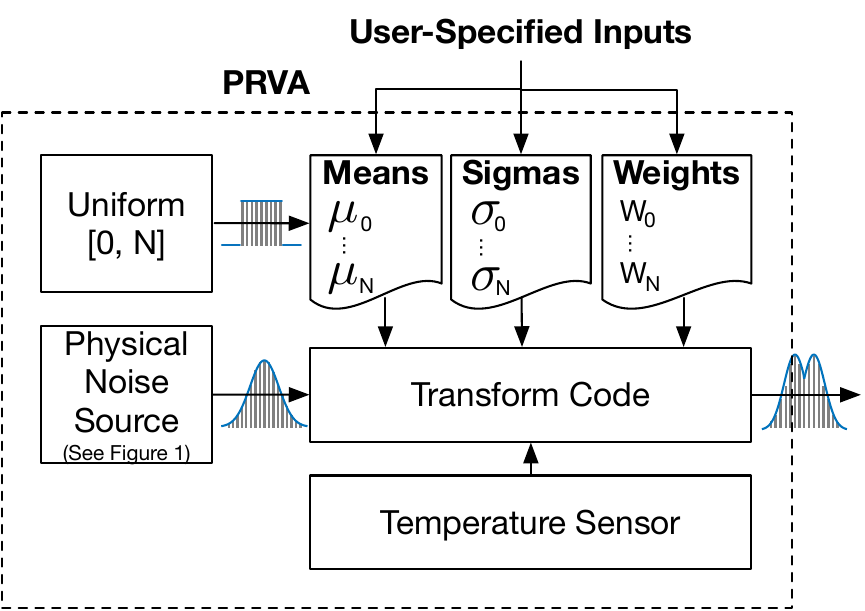}
	\caption{The programmable random variate accelerator can use a kernel density encoded as a list of means, standard deviations, and weights to randomly sample from the mixture of Gaussian distributions that makes up the kernel density.}\label{figure:kernelPRVA}
\end{figure}

Figure~\ref{figure:kernelPRVA} shows how the programmable random variate accelerator transforms the random samples in software to produce the target distribution.
The user programs the processor with three arrays containing the mean, standard deviation, and weight of each Gaussian in the kernel density. 
The processor uses a software-uniform-pseudorandom number generator~\cite{oneill2014pcg} to select a Gaussian 
to generate samples from. The transform code transforms a sample from the analog-to-digital converter to a sample from the required Gaussian.
The processor uniformly interpolates the analog-to-digital converter samples using the same uniform-pseudorandom number generator to increase their resolution from 12 to 64\,bits.
Algorithm~\ref{algorithm:onSensorInference} shows the process of transforming the generated Gaussian to the required Gaussian and then linearly interpolating with the uniform-pseudorandom-number generator.  
This process repeats each time the processor samples from the distribution. 

\begin{algorithm}
\DontPrintSemicolon
\SetAlgoLined
\SetKwInOut{Input}{input}
\Input{$\mu, \mu', \sigma, \sigma', U, X$}
\KwResult{Sample $N$ values from Gaussian random variable $X$}
$a = \frac{\sigma'}{\sigma}$\;
$b = \mu' - \mu a$\;
 \For{$i=0$ to $N$}{
	Read in random integer $x$ from the ADC\;
	$\mathrm{Sample} = \frac{x+u}{2^{64}}$\;
	$\mathrm{Samples}[i] = a*\mathrm{Sample} +b$\;
    }
RETURN Samples
\caption{Method to transform to transform $N$ samples from a distribution with a mean $\mu$ and standard deviation $\sigma$ to $N$ samples with mean $\mu'$ and standard deviation $\sigma'$.}\label{algorithm:onSensorInference}
\end{algorithm}

\section{Noise Source Temperature Dependence}

We placed the field programmable gate array development board and the Gaussian noise source inside a Binder MK56 thermal chamber. 
We set the thermal chamber to 0\,$^\circ$C and allowed the temperature to reach equilibrium for 30 minutes. 
While waiting for the temperature to reach equilibrium we had the soft processor on the field programmable gate array consistently running a program to print raw analog-to-digital converter values. 
When equilibrium was reached we ran a program to have the processor sample and print $10^6$ raw analog-to-digital converter values. 
We repeated this for ten different temperatures between 0\,$^\circ$C and 45\,$^\circ$C with a 5\,$^\circ$C temperature step.

\begin{figure*}
	\begin{subfigure}{.485\textwidth}
	  \includegraphics[width=\textwidth]{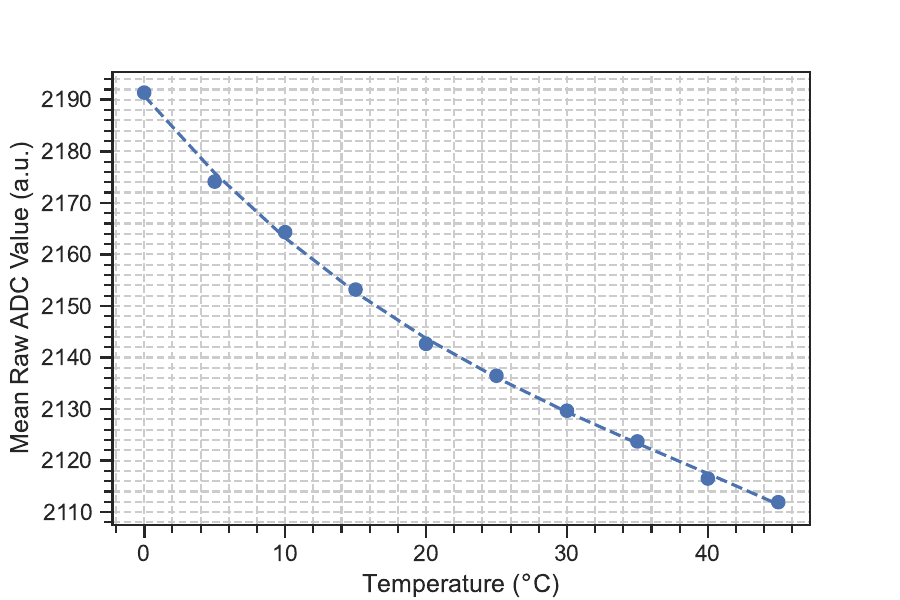}
	  \caption{Mean of raw ADC samples from the Gaussian noise source.}\label{fig:temperatureMean}
	\end{subfigure}
	\hfill
	\begin{subfigure}{.485\textwidth}
	  \includegraphics[width=\textwidth]{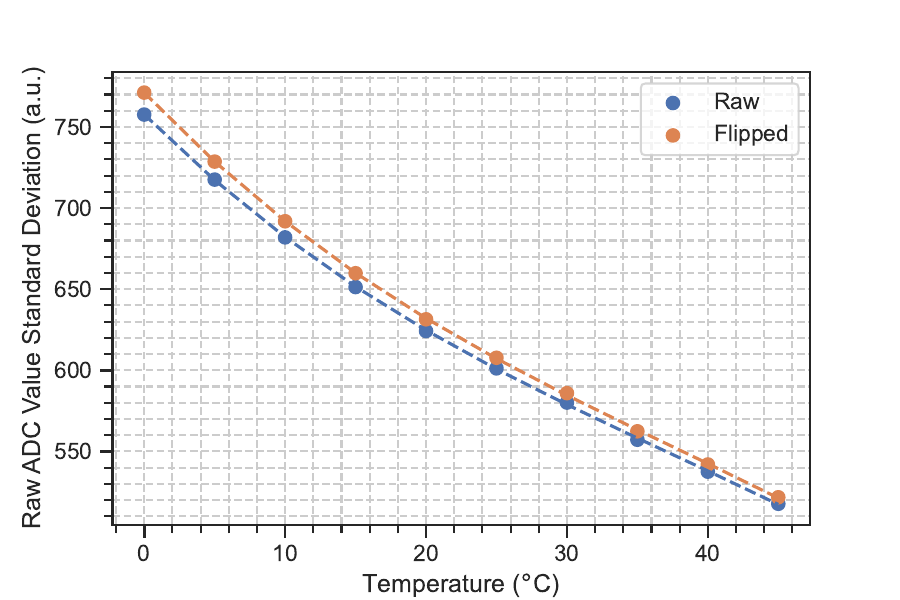}
	  \caption{Standard deviation of raw ADC samples from the Gaussian noise source.}\label{fig:temperaturestd}
	\end{subfigure}
	\caption{Mean and standard deviation of ten batches $10^6$ raw analog-to-digital converter samples over a temperature range of 0\,$^\circ$C to 45\,$^\circ$C with a 5\,$^\circ$C temperature step.
	The blue data points and curves show the mean and standard deviation of the raw ADC values. The orange curve shows the standard deviation of the raw ADC values after 50\,\% of them have been randomly subtracted from the maximum ADC output value.}\label{fig:meanSTDTemperatureDependenc}
\end{figure*}

Figure~\ref{fig:temperatureMean} shows the dependence of the mean of the raw analog-to-digital converter values on temperature.
We can eliminate this dependence by having the program that samples the analog-to-digital converter randomly subtract half of the samples from the maximum analog-to-digital converter value. 
Figure~\ref{fig:temperaturestd} shows the dependence of the standard deviation of the raw analog-to-digital converter values upon temperature. 
The curve with the legend ``Flipped'' shows that the subtraction method that removes the temperature dependence of the mean of the raw analog-to-digital converter values shifts the standard deviation upwards by a constant factor but does not remove the dependence of the standard deviation upon temperature.

\begin{figure*}
	\begin{subfigure}{.485\textwidth}
	  \includegraphics[width=\textwidth]{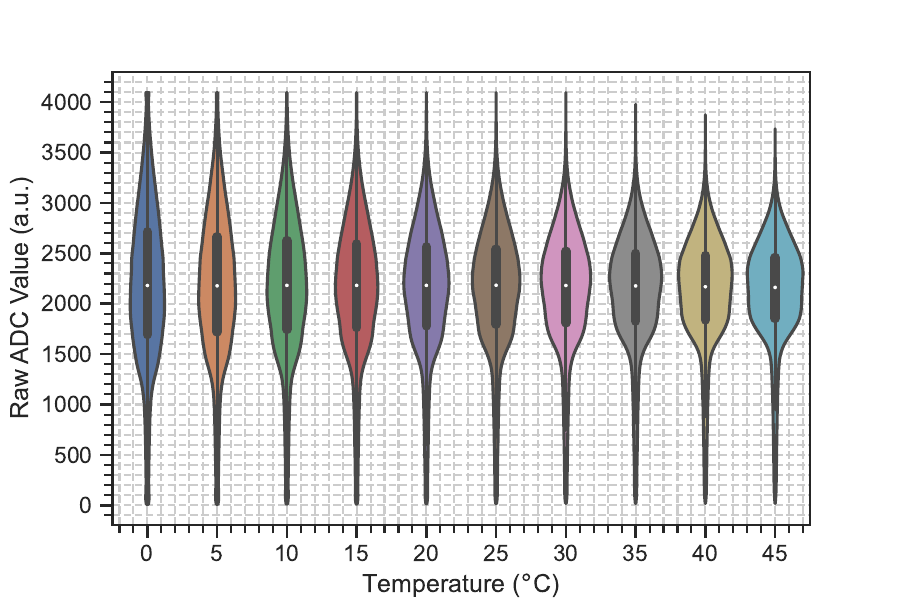}
	  \caption{Raw unprocessed ADC samples from the Gaussian noise source.}\label{fig:temperatureViolin}
	\end{subfigure}
	\hfill
	\begin{subfigure}{.485\textwidth}
	  \includegraphics[width=\textwidth]{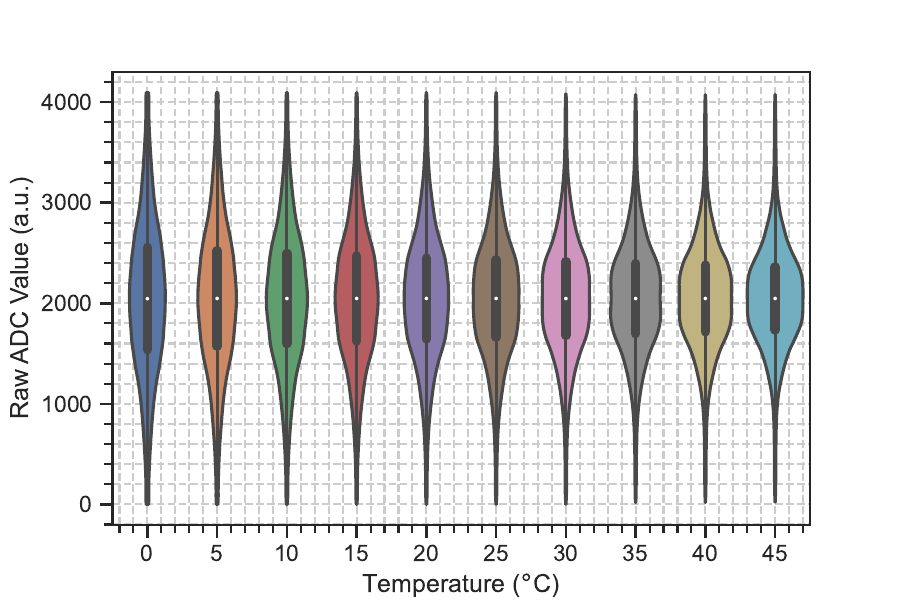}
	  \caption{Raw ADC samples randomly subtracted from 4095 with 0.5 probability.}\label{fig:flippedTemperatureViolin}
	\end{subfigure}
	\caption{Violin plot of kernel density estimates of $10^6$ ADC samples over a temperature range of 0\,$^\circ$C to 45\,$^\circ$C with a 5\,$^\circ$C temperature step.}\label{fig:distributionTemperatureDependence}
\end{figure*}

Figure~\ref{fig:temperatureViolin} shows a violin plot of the kernel density of the distribution of raw analog-to-digital converter values at each temperature.
Figure~\ref{fig:flippedTemperatureViolin} shows how randomly subtracting the raw analog-to-digital converter value from the maximum possible value creates symmetric distributions from the skewed raw samples.
This processing does not remove the temperature dependence of the standard deviation and higher-order statistics.
We should therefore measure the temperature of the Gaussian noise source and compensate for it or keep the noise source at a constant temperature when sampling it.

\section{Speed and Power Consumption Measurements}

We measured that the FemtoRV RISC-V soft processor augmented with a programmable random variate accelerator took approximately 130\,s to generate $10^9$ 64-bit random samples.
This is a sampling speed of 492\,Mb/s.
We used a Microchip ADM00921 Power Meter to measure the combined average power consumption of the FPGA development board and the low-power Gaussian noise source printed circuit board to be 1.983\,W while running the program that we used to measure the univariate Gaussian sampling speed.
We used a Keithly 2450 SourceMeter to measure the power consumption of the low-power Gaussian noise source printed circuit board to be 1.62\,mW in the on state and 32.4\,nW in the off state.

\section{Monte Carlo Program Benchmarking Study}

We benchmarked twelve C programming language applications on the LiteX-generated FemtoRV Petitbateau RISC-V processor.
The processor used a built-in 64-bit timer to measure the number of clock cycles required to run each application.
Table~\ref{Table:Applications} shows the speedup gained by using Gaussian random numbers generated by the electron tunneling noise programmable random variate instead of the GNU scientific library Gaussian and Student T random number generators.
We ran each benchmark one hundred times and found that the average speedup was $8.70 \times$ and the median speedup was $8.69 \times$.
Gaussian sampling and Gaussian mixture sampling respectively have an end-to-end speed up of 2815 $\times$ and $2899\times$ respectively if the program does not store the samples in an array.
All benchmark programs shown in Table~\ref{Table:Applications} store the results of the Monte Carlo simulation in an array.

\begin{table*}
\centering
\caption{The end-to-end speedup achieved by a programmable random variate accelerator for a range of C language benchmark applications.
		 The average speedup was $8.70 \times$ and the median speedup was $8.69 \times$.
		 The average Wasserstein distance from the reference result was $1.48\times$ and the median was $1.41\times$ that of the results produced by the GNU Scientific Library random number generators.
		 The lines column is the number of lines in the GNU scientific library implementation of the benchmark.
		 We calculate the Wasserstein distance from a $10^8$ sample reference Monte Carlo run on a workstation computer.
		 We measured the Wasserstein distance and speedup results by running 100 repeats with $10^4$ samples for each benchmark. 
		 We did this using the programmable random variate accelerator for sampling and also separately, the GNU scientific library random number generator.
		 We calculate the random sampling fraction by running each benchmark $10^4$ times and calculating the average fraction of clock cycles spent generating random samples.
		 We used the \texttt{clock()} function from the \texttt{time} C library.}\label{Table:Applications}
\begin{tabular}{ccccccc}
\toprule
\multirow{2}{*}{\textbf{Application}}                               & \multirow{2}{*}{\textbf{\begin{tabular}[c]{@{}c@{}}Wasserstein Distance \\ Ratio PRVA / GSL \end{tabular}}} & \multirow{2}{*}{\textbf{\begin{tabular}[c]{@{}c@{}}Random Sampling \\ Fraction (\%)\end{tabular}}} & \multirow{2}{*}{\textbf{\begin{tabular}[c]{@{}c@{}}End-to-End \\ Speed Up ($\times$)\end{tabular}}} & \multirow{2}{*}{\textbf{\begin{tabular}[c]{@{}c@{}}Number \\ of Lines \end{tabular}}} & \multirow{2}{*}{\textbf{\begin{tabular}[c]{@{}c@{}}Sampling \\ Distribution \end{tabular}}} & \multirow{2}{*}{\textbf{Source}}                    \\
                                                                    &                                                                                                             &                                                                                                    &                                                                                                     &                                                                                       &                                                                                             &                                                     \\
\midrule
\rowcolor{a} Gaussian Sampling									    & 1.98                                                                                                        & 98.8                                                                                               & 9.36                                                                                                & 80                                                                                    & Gaussian                                                                                    & This Work                                           \\
\rowcolor{b} Gaussian Mixture									    & 1.17                                                                                                        & 97.5                                                                                               & 6.89                                                                                                & 80                                                                                    & Mixture                                                                                     & This Work                                           \\
\rowcolor{a} Addition                                               & 1.12                                                                                                        & 92.1                                                                                               & 9.31                                                                                                & 87                                                                                    & Gaussian                                                                                    & \cite{signaloidAddition, tsoutsouras2021laplace}    \\
\rowcolor{b} Divide                                                 & 1.51                                                                                                        & 92.1                                                                                               & 8.59                                                                                                & 87                                                                                    & Gaussian                                                                                    & \cite{signaloidDivision, tsoutsouras2021laplace}    \\
\rowcolor{a} Multiply                                               & 1.61                                                                                                        & 92.4                                                                                               & 8.78                                                                                                & 87                                                                                    & Gaussian                                                                                    & \cite{signaloidMultiply, tsoutsouras2021laplace}    \\
\rowcolor{b} Subtract                                               & 1.21                                                                                                        & 92.2                                                                                               & 10.24                                                                                               & 87                                                                                    & Gaussian                                                                                    & \cite{signaloidSubtraction, tsoutsouras2021laplace} \\
\rowcolor{a} Schlieren                                              & 1.26                                                                                                        & 91.5                                                                                               & 8.83                                                                                                & 87                                                                                    & Gaussian                                                                                    & \cite{signaloidSchlieren, tsoutsouras2021laplace}   \\
\rowcolor{b} NIST-UM Dynamic Viscosity                              & 1.84                                                                                                        & 96.0                                                                                               & 6.88                                                                                                & 90                                                                                    & Gaussian                                                                                    & \cite{signaloidViscosity, tsoutsouras2021laplace}   \\
\rowcolor{a} NIST-UM Thermal Expansion Coefficient				    & 1.30      			                                                                                      & 98.3                                                                                               & 25.24									                                                             & 84			                                                                         & Student-T                                                                                   & \cite{signaloidThermal, tsoutsouras2021laplace}      \\
\rowcolor{b} Medical Covid-19 R0                                    & 1.09                                                                                                        & 82.5                                                                                               & 5.40                                                                                                & 143                                                                                   & Mixture                                                                                     & \cite{signaloidCovid}                               \\
\rowcolor{a} Geometric Brownian Motion                              & 1.72                                                                                                        & 69.3                                                                                               & 2.35                                                                                                & 102                                                                                   & Gaussian                                                                                    & \cite{oosterlee2019mathematical}                    \\
\rowcolor{b} Black Scholes Monte Carlo Pricing                      & 1.93                                                                                                        & 71.9                                                                                               & 2.57                                                                                                & 98                                                                                    & Gaussian                                                                                    & \cite{armstrong2017c++}                             \\

\bottomrule
\end{tabular}
\end{table*}

\section{Related Research}\label{section:relatedwork}

\subsection{Programmable Random Variate Accelerators}

Table~\ref{Table:gaussian} shows a summary of state-of-the-art programmable random variate accelerators.
As far as we are aware this article presents the fastest and most efficient electronic noise programmable random variate accelerator that can generate samples from any univariate non-uniform probability distribution. 
Three approaches from Table~\ref{Table:gaussian} record a lower power consumption than our approach but are inferior in other aspects. 
The power measurement for the memristor-based approach~\cite{jiang2017novel} does not include the power consumption of any of the circuitry required to sample the device or bias it to the correct voltage. 
The power quoted for the resonance energy transfer approach~\cite{zhang2018architecting} is a back-of-the-envelope estimate and not experimentally measured. 
The implementation we present in this paper is approximately $35,652 \times$ and $72 \times$ faster than programmable random variate accelerators we presented in previous publications~\cite{meech2020efficient, myInvention}. 
As the sampling speed is mainly limited by the analog-to-digital converter sampling rate, using LiteX to deploy the design on FPGAs with a higher analog-to-digital converter sampling rate will lead to further speed increases.

\begin{table*}
    \centering
    \caption{Comparison of state-of-the-art programmable random variate accelerator (PRVA) methods~\cite{meech2020efficient}.}
    \begin{tabular}{ccccccc}
    \toprule
    \rowcolor{b} \bf{Source} & \bf{Speed} & \bf{Efficiency} & \bf{Power} & \bf{Distribution(s)} & \bf{Programmable}        & \bf{Publication}                       \\
    \midrule
    \rowcolor{a} Central Processing Unit            & 890 Mb/s   & 3.17\,Mb/J      & 281\,W     & Gaussian     & Yes       & \cite{thomas2009comparison}, 2009      \\
    \rowcolor{b} Graphics Processing Unit           & 12.9 Gb/s  & 108\,Mb/J       & 119\,W     & Gaussian     & Yes       & \cite{thomas2009comparison}, 2009      \\
    \rowcolor{a} Massively Parallel Processor Array & 860 Mb/s   & 403\,Mb/J       & 2.13\,W    & Gaussian     & Yes       & \cite{thomas2009comparison}, 2009      \\
    \rowcolor{b} Field Programmable Gate Array      & 12.1 Gb/s  & 645\,Mb/J       & 18.8\,W    & Gaussian     & Yes       & \cite{thomas2009comparison}, 2009      \\
    \hline
    \rowcolor{a} Memristor                          & 6000 b/s   & 120\,Gb/J       & 50.0\,nW   & Unnamed      & No        & \cite{jiang2017novel}, 2017            \\
    \rowcolor{b} Photo Detector                     & 1.77 Gb/s  & -               & -          & Gaussian     & No        & \cite{marangon2017source}, 2017        \\
    \rowcolor{a} Resonance Energy Transfer          & 2.89 Gb/s  & 578\,Gb/J       & 5.00\,mW   & Exponential  & Yes       & \cite{zhang2018architecting}, 2018     \\
    \rowcolor{b} Photo Diode                        & 17.4 Gb/s  & -               & -          & Husumi       & No        & \cite{avesani2018source}, 2018         \\
    \rowcolor{a} Photo Diode                        & 66.0 Mb/s  & -               & -          & Programmable & Yes       & \cite{nguyen2018programmable}, 2018    \\
    \rowcolor{b} Photo Diode                        & 320 Mb/s   & -               & -          & Exponential  & No        & \cite{tomasi2018model}, 2018           \\
    \rowcolor{a} Field Programmable Gate Array      & 6.40 Gb/s  & -               & -          & Gaussian     & Yes       & \cite{hu2019gaussian}, 2019            \\
    \rowcolor{b} Photo Diode                        & 8.25 Gb/s  & -               & -          & Gaussian     & No        & \cite{guo2019parallel}, 2019           \\
    \rowcolor{a} Electronic Noise                   & 13.8 kb/s  & 209\,kb/J       & 66.0\,mW   & Gaussian     & Yes       & \cite{meech2020efficient}, 2020        \\
	\rowcolor{b} Pseudorandom                       & -          & -               & 78.9\,mW   & Programmable & Yes       & \cite{bashizade2021accelerating}, 2021 \\  
	\rowcolor{a} Electronic Noise                   & 6.82 Mb/s  & 13.4\,Mb/J      & 484\,mW    & Programmable & Yes       & \cite{myInvention}, 2022    			\\
	\rowcolor{y} Electronic Noise                   & 492\,Mb/s  & 248\,Mb/J       & 1.98\,W    & Programmable & Yes       & This work     					        \\
    \bottomrule
    \end{tabular}
    \normalsize\label{Table:gaussian}
\end{table*}

\subsection{Thermodynamic Computing}
 
Thermodynamic computing~\cite{conte2019thermodynamic, todd2021vision} uses hardware capable of sampling from an arbitrary two-dimensional Gaussian for solving a linear system of equations, estimating the inverse of a matrix, solving the Lyapunov equation, and estimating the determinant of a matrix~\cite{coles2023thermodynamic, aifer2023thermodynamic, duffield2023thermodynamic, aifer2024error}. 
A blog post describing thermodynamic computing hardware for Gaussian sampling reports a speedup of approximately $13 \times$ and an energy saving of approximately $25 \times$.
Detailed systems evaluations explaining their hardware architecture and how it achieves this are not available~\cite{normal-speedup}. 

\subsection{Tuned Stochastic Probability Trees}

Prior work to design and build stochastic magnetic tunnel junction devices to implement a computing device that can perform single random bit sampling with a programmable bias exists~\cite{cardwell2022probabilistic, 9998481, misra2023probabilistic, theilman2022stochastic, rehm2023stochastic, 10272252, rehm2023temperatureresilient, aimone2023synaptic, wolpert2023stochastic, maicke2023magnetic}.
This approach is very different from our approach as it builds up probability distributions using many single programmable random bits. 

\section*{Conclusion}

This article presented an electron tunneling noise programmable random variate accelerator for accelerating the sampling stage of Monte Carlo simulations.
We used the LiteX framework to generate a Petitbateau FemtoRV RISC-V instruction set soft processor and deploy it on a Digilent Arty-100T FPGA development board.
The RISC-V soft processor augmented with our programmable random variate accelerator achieves an average speedup of $8.70 \times$ and a median speedup of $8.69 \times$ for a suite of twelve different benchmark applications when compared to GNU Scientific Library software random number generation. 
These speedups are achievable because the benchmarks spend an average of 90.0\,\% of their execution time generating random samples. 
The results of the Monte Carlo benchmark programs run over the programmable random variate accelerator have an average Wasserstein distance of $1.48\times$ and a median Wasserstein distance of $1.41\times$ that of the results produced by the GNU Scientific Library random number generators.
The soft processor samples the electron tunneling noise source using the hardened XADC block in the FPGA.
The flexibility of the LiteX framework allows for the deployment of any LiteX-supported soft processor with an electron tunneling noise programmable random variate accelerator on any LiteX-supported development board that contains an FPGA with an XADC.

\acknow{We thank Florent Kermarrec and all the contributors to LiteX for creating the open-source tool we leveraged to create a portable implementation of our architecture~\cite{kermarrec2020litex}.
	    We thank Bruno Levy for creating LiteOS and for extensive help getting LiteOS working to load elf program files into the LiteX-generated soft processor on an FPGA~\cite{liteos}.
	    In addition, we thank Andrew "Bunnie" Huang for extensive help and advice in getting the XADC working with the LiteX framework.}

\showacknow 

\vspace{0.25in}

\section*{References}
\bibliography{working-document}

\onecolumn
\appendix

\twocolumn

\end{document}